\documentclass[preprint,12pt]{elsarticle}                                                                                          
\usepackage{epsfig}                                                                                               
\usepackage{amssymb}    
\usepackage{rotating}
\usepackage{tikz}                                                                                          
                                                                                                                  
\journal{Physics Letters B}                                                                                       
                                                                                                                  
\begin{document}                                                                                                  
                                                                                                                  
\begin{frontmatter}                                                                                               
                                                                                                                  
\title{First determination of the one-proton induced \\ Non-Mesonic Weak Decay width \\ of $p$-shell 
$\Lambda$-Hypernuclei}     
\author{The FINUDA Collaboration}                                                                   
\author[a,b]{M.~Agnello}                                         
\author[c]{L.~Benussi} \author[c]{M.~Bertani} \author[d]{H.C.~Bhang}                                           
\author[e,f]{G.~Bonomi} \author[g,b]{E.~Botta \corref{cor1}}  
\author[g,b]{T.~Bressani} \author[b]{S.~Bufalino}                                   
\author[b]{D.~Calvo} \author[h,i]{P.~Camerini}                                          
\author[k,l]{B.~Dalena  \fnref{foot1}}
\author[g,b]{F.~De Mori} \author[k,l]{G.~D'Erasmo}                                     
\author[b]{A.~Feliciello} \author[b]{A.~Filippi}                                       
\author[m]{H.~Fujioka}                                         
\author[c]{P.~Gianotti} \author[i]{N.~Grion}                                                                                           
\author[c]{V.~Lucherini} \author[g,b]{S.~Marcello}                                     
\author[m]{T.~Nagae} \author[p]{H.~Outa} 
\author[k]{V.~Paticchio} \author[i]{S.~Piano} \author[h,i]{R.~Rui} \author[k,l]{G.~Simonetti} 
\author[e,f]{A.~Zenoni}                                                                                                                                                                                           
                                                 
\address[a]{DISAT, Politecnico di Torino, corso Duca degli Abruzzi 24, Torino, Italy}                                                                                        
\address[b]{INFN Sezione di Torino, via P. Giuria 1, Torino, Italy}                                               
\address[c]{Laboratori Nazionali di Frascati dell'INFN, via. E. Fermi, 40, Frascati, Italy}                                                                                                  
\address[d]{Department of Physics, Seoul National University, 151-742 Seoul, South Korea}
\address[e]{Dipartimento di Ingegneria Meccanica e Industriale, \\ Universit\`a di Brescia, via Branze 38,                                     
Brescia, Italy}                                                                                                   
\address[f]{INFN Sezione di Pavia, via Bassi 6, Pavia, Italy}                                                     
\address[g]{Dipartimento di Fisica, Universit\`a di Torino, via P. Giuria 1, Torino, Italy}                                                                                   
\address[h]{Dipartimento di Fisica, Universit\`a di Trieste, via Valerio 2, Trieste, Italy}                                                                                                   
\address[i]{INFN Sezione di Trieste, via Valerio 2, Trieste, Italy}                                               
\address[k]{INFN Sezione di Bari, via Amendola 173, Bari, Italy}                                                                                                
\address[l]{Dipartimento di Fisica Universit\`a di Bari, via Amendola 173, Bari, Italy}                       
\address[m]{Department of Physics, Kyoto University, Sakyo-ku, Kyoto Japan} 
%%\address[n]{Department of Physics, Shahid Behesty University, 19834 Teheran, Iran}                                                                                                             
%%\address[o]{INAF-IFSI, Sezione di Torino, Corso Fiume 4, Torino, Italy}                                           
\address[p]{RIKEN, Wako, Saitama 351-0198, Japan}                                                               
                                                                            
\cortext[cor1]{Corresponding author. E-mail address: botta@to.infn.it}                                                                                                                                                                                                            
\fntext[foot1]{Now at CEA/SACLAY, DSM/Irfu/SACM F-91191 Gif-sur-Yvette France}
\begin{abstract} 
Previous studies of proton and neutron spectra from Non-Mesonic Weak Decay of eight $\Lambda$-Hypernuclei ($A=5 \div 16$) have been revisited. New values of the ratio of the two-nucleon 
and the one-proton induced decay widths, $\Gamma_{2N} / \Gamma_{p}$, are obtained from single 
proton spectra, \mbox{$\Gamma_{2N} / \Gamma_{p} = 0.50 \pm 0.24$}, and from neutron and proton coincidence spectra, \mbox{$\Gamma_{2N} / \Gamma_{p} = 0.36\pm {0.14_{stat}}^{+0.05_{sys}}_{{-{0.04_{sys}}}}$}, in full agreement with previously published ones. With these values, a method is developed to extract the one-proton induced decay width in units of the free $\Lambda$ decay width, $\Gamma_{p}/\Gamma_{\Lambda}$, without resorting to Intra Nuclear Cascade models but by exploiting only experimental data, under the assumption of a linear dependence on $A$ of the Final State Interaction contribution.  This is the first systematic determination ever done and it agrees within the errors with recent theoretical calculations.
\end{abstract}                                                                                                                                      

                                                                                                                  
\end{frontmatter}                                                                                                 
                                                                                                                  
\section{Introduction}           
$\Lambda$-Hypernuclei (Hypernuclei in the following) decay through Weak Interaction to non-strange nuclear systems following two modes, the mesonic (MWD) and the non-mesonic (NMWD) one. 
The MWD is further split into two branches corresponding to the decay modes of the $\Lambda$ in 
free space:
\begin{equation}
^{A}_{\Lambda}Z \rightarrow ^{A}(Z+1)  + \pi^{-}       \quad(\Gamma_{\pi^{-}})                                       
\label{mwd1}
\end{equation}
\begin{equation}
^{A}_{\Lambda}Z \rightarrow ^{A}Z + \pi^{0}         \quad(\Gamma_{\pi^{0}}).
\label{mwd2}
\end{equation}
$^{A}_{\Lambda}Z$ indicates the Hypernucleus with mass number $A$ and atomic number $Z$, $^{A}(Z+1)$ and $^{A}Z$ the residual nuclear system, usually the daughter nucleus in its ground state, and the $\Gamma$'s stand for the decay widths. Since the momentum released to the nucleon in MWD 
(p$\sim$100 MeV/c, Q$_{MWD}\sim$37 MeV) is much lower than the Fermi momentum, 
the MWD is strongly suppressed by the Pauli exclusion principle in all but the lightest Hypernuclei.
In NMWD the Hypernucleus decays through Weak Interaction involving the constituent $\Lambda$ and one or more core nucleons. The importance of such processes was pointed out for the first time in \cite{ches}. 
If the pion emitted in the weak vertex $\Lambda \rightarrow \pi N$ is virtual, then 
it can be absorbed by the nuclear medium giving origin to:
\begin{equation}
^{A}_{\Lambda}Z  \rightarrow ^{A-2}(Z-1) + n + p \quad(\Gamma_{p})~,
 \label{gammap} 
\end{equation}
\begin{equation}
^{A}_{\Lambda}Z \rightarrow ^{A-2}Z + n + n \quad(\Gamma_{n})~,
\label{gamman} 
\end{equation}
\begin{equation}
^{A}_{\Lambda}Z \rightarrow ^{A-3}(Z-1) + n + n + p \quad(\Gamma_{2N})~.
\label{gamma2} 
\end{equation}
The processes (\ref{gammap}) and (\ref{gamman}) are globally indicated as one-nucleon induced decays 
(one-proton (\ref{gammap}), one-neutron (\ref{gamman})) while (\ref{gamma2}) as two-nucleon induced decay. By neglecting $\Lambda$ weak interactions with nuclear clusters of more than two nucleons, 
the total NMWD width is: 
\begin{equation}
\Gamma _{NMWD} = \Gamma_{p} + \Gamma_{n} + \Gamma_{2N}~.
\label{gammatot}
\end{equation}
The two-nucleon induced mechanism (\ref{gamma2}) was first suggested in \cite{albe} and interpreted by assuming that the virtual pion from the weak vertex is absorbed by a pair of nucleons ($np$, $pp$ or $nn$), correlated by the strong interaction. In (\ref{gamma2}) we have indicated for simplicity only the most probable process involving $np$ pairs. Note that the NMWD can also be mediated by the exchange of mesons more massive than the pion.

The NMWD mode is possible only in nuclei; the Q-value of the elementary weak reactions driving the decays (\ref{gammap}), (\ref{gamman}) and (\ref{gamma2}) is high enough (Q$_{NMWD}\sim$175 MeV) to avoid any Pauli blocking effect and the final nucleons thus have a large probability to escape from the nucleus. Indeed it is expected that NMWD dominates over MWD for all but the $s$-shell Hypernuclei and only for very light systems the two decay modes are expected to be competitive.
The total decay width of an Hypernucleus, $\Gamma_{T}$, is thus given by:  
\begin{equation}
\Gamma_{T}  = \Gamma_{\pi^{-}} + \Gamma_{\pi^{0}} + \Gamma_{p} + \Gamma_{n} + \Gamma_{2N}~.
\label{gammatot2}
\end{equation}
In order to compare data from different Hypernuclei, partial $\Gamma$'s are usually given in units of  $\Gamma_{\Lambda}$, the total decay width of the free $\Lambda$.   

The NMWD of Hypernuclei has been scarcely studied up to a few years ago. Experimentally it is not only necessary to produce and to identify Hypernuclei in their ground state by means of a performing magnetic spectrometer, but also to detect in coincidence the nucleons emitted in (\ref{gammap}), (\ref{gamman}) and (\ref{gamma2}) and to measure their energy. Furthermore, there is a big difficulty in  extracting the observables related to the above mentioned weak processes due to the 
strong distorsion introduced by Final State Interaction (FSI) on the spectra of the nucleons emitted in the elementary processes corresponding to (\ref{gammap}), (\ref{gamman}), (\ref{gamma2}). The information on the initial bare momenta may be completely lost, and their contributions can be mixed, with possible additional quantum-mechanical interference effects \cite{garb1}. 

Following the first pioneering experiments \cite{szym,noumi}, the SKS Collaboration measured at the 12 GeV KEK PS the spectra of protons and neutrons from the NMWD 
of $^{5}_{\Lambda}$He and $^{12}_{\Lambda}$C produced by the ($\pi^{+}$, $K^{+}$) reaction at 1.05 GeV/c 
\cite{okadanmwd}. 
At the DA$\Phi$NE  ($e^{+}$, $e^{-}$) collider the FINUDA Collaboration measured the spectra of protons and neutrons from $^{5}_{\Lambda}$He and from 7 $p$-shell Hypernuclei. 
A full account of all the papers on the subject may be found in \cite{epja}. 
Many theoretical papers have also been produced, motivated by the strong interest and discovery potential of the study of NMWD of Hypernuclei; they are listed \cite{epja} as well.

\section{A revisited analysis of the proton spectra from FINUDA}

In an early paper \cite{npa804} proton spectra from NMWD of $^{5}_{\Lambda}$He, $^{7}_{\Lambda}$Li and $^{12}_{\Lambda}$C, measured by FINUDA, were presented and discussed. 

As a second step, proton spectra from NMWD of $^{9}_{\Lambda}$Be, $^{11}_{\Lambda}$B, $^{13}_{\Lambda}$C, $^{15}_{\Lambda}$N and $^{16}_{\Lambda}$O were produced and analyzed \cite{plb685}; the experimental energy resolution was $\Delta E / E = 2 \%$ at 80 MeV.
\begin{sidewaysfigure}[htbp]
\begin{center}
\includegraphics[width=200mm]{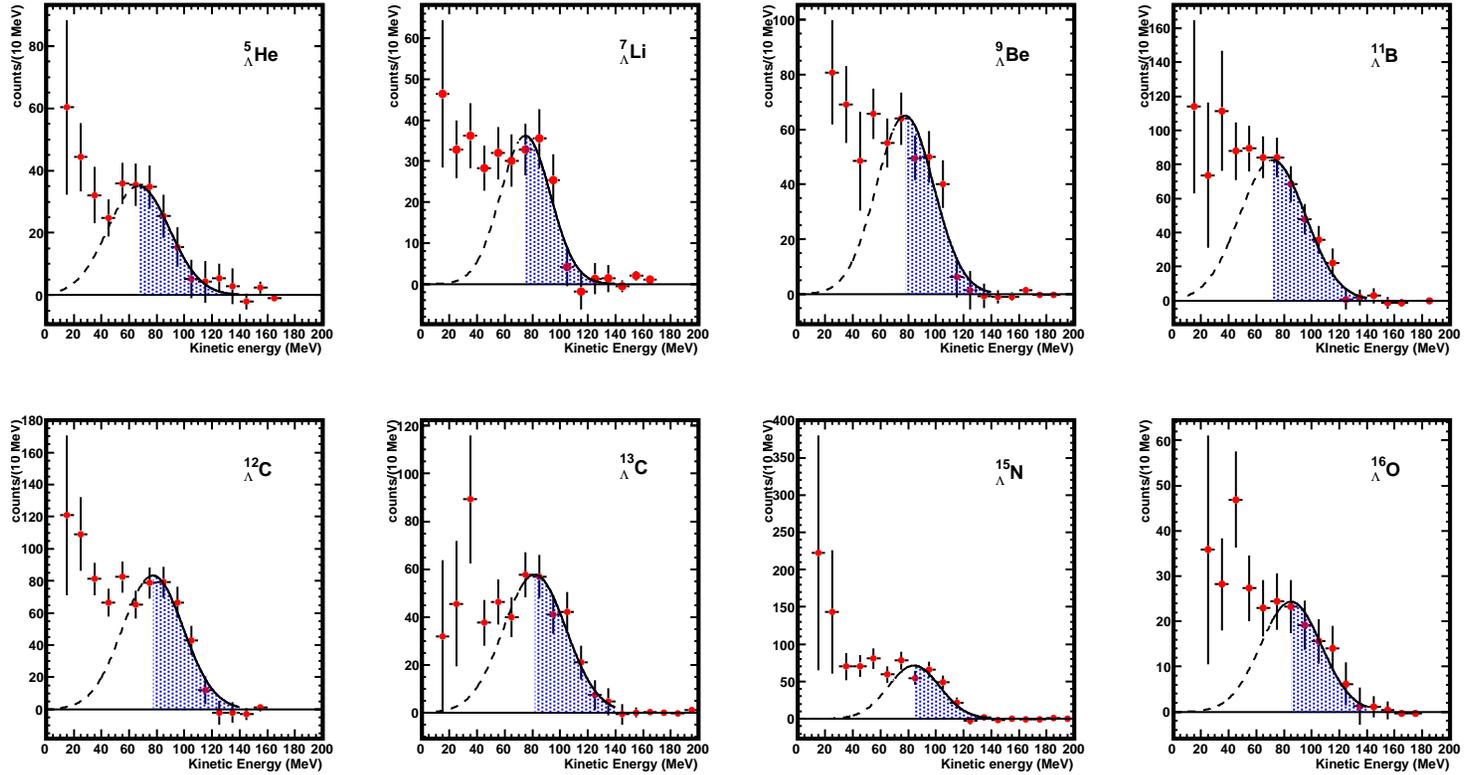}
\caption{(Color online) Proton kinetic energy spectra from the NMWD of (from left to right up and down rows): ${\mathrm{^{5}_{\Lambda}He}}$,  ${\mathrm{^{7}_{\Lambda}Li}}$, ${\mathrm{^{9}_{\Lambda}Be}}$,  ${\mathrm{^{11}_{\Lambda}B}}$,  ${\mathrm{^{12}_{\Lambda}C}}$,  ${\mathrm{^{13}_{\Lambda}C}}$,  ${\mathrm{^{15}_{\Lambda}N}}$ and 
 ${\mathrm{^{16}_{\Lambda}O}}$. The curves represent the new analysis gaussian fits to the spectra: the solid 
line part indicates the actual fit region, the dashed part indicates the one proton induced NMWD contribution  
to the lower energy spectrum part. The blue filled area is the higher energy half gaussian area, where the two-nucleon induced NMWD is negligible.} 
\label{pspectra}
\end{center}
\end{sidewaysfigure}
Figure~\ref{pspectra} shows all the experimental spectra; they are up to now a unique data bank for $p$-shell Hypernuclei with $A=5 \div 16$, from which several interesting considerations and conclusions were drawn. 

In \cite{plb685} we developed a method for disentangling the contributions from the decay (\ref{gamma2}) without using IntraNuclear Cascade (INC) calculations, as in \cite{kim,bauer}. 
The first step was to fit the eight experimental spectra of Figure~\ref{pspectra} above 80 MeV 
($\sim Q_{NMWD}/2$) to Gaussians with free central values, widths and areas. In Table \ref{table1} the values of the Gaussian’s centers, $\mu_{0}$, with their statistical error are reported.

\begin{table}[h] 
\begin{center} 
\begin{tabular}{|c|c|c|c|c|c|} 
\hline 
 & $\rho$ & $\mu_{0}$ & $\mu_{1}$ & $\sigma_{1}$  & $\mu_{2}$ \\ 
                     & ( MeV) &  (MeV)  & (MeV) & (MeV) & (MeV) \\  
\hline 
${\mathrm{^{5}_{\Lambda}He}}$ &76.65  & 68.5$\pm$4.1 & 66.9$\pm$11.8 &  22.3$\pm$9.9 & 65.0$\pm$16.9\\ 
\hline
${\mathrm{^{7}_{\Lambda}Li}}$ & 82.99 & 76.7$\pm$5.2 & 74.9$\pm$3.8 & 18.0$\pm$2.1 & 77.7$\pm$2.9 \\ 
\hline
${\mathrm{^{9}_{\Lambda}Be}}$ & 76.48 & 78.2$\pm$6.2 & 77.7$\pm$9.1 & 20.8$\pm$10.8 & 77.3$\pm$3.8 \\ 
\hline 
${\mathrm{^{11}_{\Lambda}B}}$ & 79.72 & 75.1$\pm$5.0 & 71.7$\pm$10.8 & 23.8$\pm$5.5 & 70.0$\pm$6.3  \\ 
\hline 
${\mathrm{^{12}_{\Lambda}C}}$ & 78.36 & 80.2$\pm$2.1 & 77.3$\pm$2.9 & 22.0$\pm$2.1 & 79.9$\pm$2.2 \\ 
\hline 
${\mathrm{^{13}_{\Lambda}C}}$ & 74.44 & 83.9$\pm$12.8 & 81.6$\pm$5.8 & 22.6$\pm$3.5 & 82.8$\pm$3.1 \\  
\hline 
${\mathrm{^{15}_{\Lambda}N}}$ & 77.55 & 88.1$\pm$6.2 & 84.2$\pm$4.5 & 18.6$\pm$2.8 & 80.6$\pm$3.3 \\  
\hline 
${\mathrm{^{16}_{\Lambda}O}}$ & 78.25 & 93.1$\pm$6.2 & 85.0$\pm$6.8 & 21.9$\pm$3.5 & 81.0$\pm$5.7\\  
\hline 
\end{tabular} 
\caption{Kinematics and Gaussian fit parameters. First column: hypernucleus; second column: proton kinetic energy, $\rho$, from a 2-body kinematics of one proton induced NMWD, with no daughter nucleus recoil (see text for more details); third column: gaussian fit mean value from \cite{plb685}, $\mu_{0}$; fourth column: present analysis gaussian fit mean value, $\mu_{1}$; fifth column:  present analysis gaussian fit standard deviation, $\sigma_{1}$; sixth column: gaussian fit mean value starting form 60 MeV, $\mu_{2}$. Statistical errors only are quoted.} 
\label{table1} 
\end{center} 
\end{table}

Recently it was outlined in \cite{npa914} that the values of $\mu_{0}$ for $^{13}_{\Lambda}$C,  $^{15}_{\Lambda}$N and $^{16}_{\Lambda}$O were significantly larger than those calculated following the relativistic kinematics  with the exact Q-values for the considered eight decays (\ref{gammap}), in the hypothesis of a back-to-back emission of the proton-neutron pair with no recoil of the residual nucleus in its ground state; they are reported in the second column of Table 1 and will be labelled $\rho$ in the following. 
The reduced $\chi^{2}$ of $\mu_{0}$ values with respect to $\rho$ ones, $\chi^{2}/ ndf= \sum_{i=1}^{8}(\mu_{0i}-\rho_{i})^{2}/16 \sigma^{2}_{\mu_{0i}}$, was 1.88. 

In the present work, we check whether better results can be obtained by shifting down the lower edge of the fitting interval of the experimental 
spectra.  
The considered $\chi^{2}$ is minimised in the hypothesis of no recoil of the residual nucleus. In general, the decay  
happens obeying to both energy and momentum conservation and the kinetic energy of the daughter nucleus is  negligible only for higher masses: 
indeed, in \cite{npa914}, it was found that for the lighter nuclei,  $^{5}_{\Lambda}$He and $^{7}_{\Lambda}$Li, a recoil momentum of $\sim$200 MeV/c allows to reproduce the obtained $\mu_{0}$ values while the corresponding $\rho$ values are not compatible within the errors.  
Nevertheless, we chose to minimize this particular $\chi^{2}$ to investigate the higher mass region keeping in mind that in the lower mass region 
only loosely bound or not bound light daughter nuclei are produced. 

For 70 MeV we find a reduced $\chi^{2}$ of 1.33, for 60 MeV of 1.86  and for 50 MeV of 3.61. We conclude that the most appropriate choice 
is to fit all experimental spectra starting from 70 MeV.  
We discard starting from 50 MeV and we consider the options of starting from 60 and 80 MeV to estimate systematic errors: their value is 
$\leq 3.5 \%$ up to $^{13}_{\Lambda}$C and increases to $4.5 \%$ for $^{15}_{\Lambda}$N and to $7.1 \%$ for $^{16}_{\Lambda}$O, where 
it is comparable to the statistical error.  
The  new Gaussians central values, $\mu_{1}$, are reported in the fourth column of Table 1, whereas in the fifth column the values of  the corresponding widths and in the sixth column the Gaussian central values for fit from 60 MeV, $\mu_{2}$, are reported. 
The quoted errors are statistical only. 

For sake of completeness, we observe that it is not useful to try to fit the proton spectra starting from values higher than 80 MeV because only 
very few points would be used and bigger errors would be obtained on the Gaussians parameters. In addition, in order to obtain satisfying fits 
it is necessary to constrain the Gaussians central values in quite small ranges, while it is not necessary to constrain them in the considered fit intervals.

We notice that the widths found for $^{5}_{\Lambda}$He and $^{12}_{\Lambda}$C are consistent with those evaluated theoretically as due to the Fermi motion \cite{garb1}. The new fitting Gaussians are represented by the solid lines in Figure~\ref{pspectra}. 

The most relevant issue from this revisited analysis is that new values for the areas of the upper half of the fitting Gaussians are evaluated, with impact on the related physics items that are discussed in the following. 

\section{A refined determination of $\Gamma_{2N} / \Gamma_{NMWD}$}

In \cite{plb685} a technique was devised to disentangle the contribution coming from the 2N induced 
decays (\ref{gamma2}) from the one-proton induced decays affected by FSI by exploiting the systematics in the mass range A=5$\div$16. Each spectrum of Figure~1 was divided into two parts, one below the value $\mu_{0}$, with area $A_{low}$, the other above, with area $A_{high}$. Since we find that the new curves, centered at $\mu_{1}$, 
provide a better description of the experimental spectra, we calculate the new values of $A_{low}$ and $A_{high}$ (blue filled areas in Figure~\ref{pspectra}). 
New values of the ratio  $R=A_{low}/(A_{low}+A_{high})$ are found and 
we repeat then exactly the same procedure as in \cite{plb685}.
Finally we find the new values:
\begin{equation}
\Gamma_{2N}/\Gamma_{p}  =  0.50 \pm 0.24_{stat} \pm0.04_{sys} 
\label{res_p1}
\end{equation}
and
\begin{equation}
\Gamma_{2N}/\Gamma_{NMWD}  =  0.25 \pm 0.12_{stat} \pm0.02_{sys} 
\label{res_p}
\end{equation}
by using the value $\Gamma_{n}/\Gamma_{p}$= (0.48$\pm$0.08), weighted average (w.a. from now on)  taken from the data in 
\cite{bhang2}. We recall that the main assumptions in the above procedure are a linear dependence of the 
FSI contribution on $A$ and the constancy of both $\Gamma_{2N}/\Gamma_{NMWD}$ and $\Gamma_{n}/\Gamma_{p}$ for Hypernuclei in the range $A=5\div16$ under consideration, as discussed in 
\cite{albe2}. We remark that the new values are fully consistent with the previous ones
($\Gamma_{2N}/\Gamma_{p} = 0.43 \pm 0.25$, $\Gamma_{2N}/\Gamma_{NMWD} = 0.24 \pm 0.10$); the smaller relative error on $\Gamma_{2N}/\Gamma_{p}$ is due to the larger $A_{high}$ integral obtained with the new fits from 70 MeV, while the error on $\Gamma_{2N}/\Gamma_{NMWD}$ is dominated 
by the error on the w.a. from \cite{bhang2}.
The systematic error quoted in (\ref{res_p1}) refers to the maximum difference obtained considering the fits 
from 60 MeV, $\Gamma_{2N}/\Gamma_{p} = 0.42 \pm 0.21$, and 80 MeV, $\Gamma_{2N}/\Gamma_{p} = 0.43 \pm 0.25$ \cite{plb685} too; 
in (\ref{res_p}) the systematic error is calculated by propagating the previous one. 

In a second approach  \cite{plb701} we determined $\Gamma_{2N}/\Gamma_{NMWD}$ by considering 
both protons and neutrons emitted in coincidence with the $\pi^{-}$ from the formation reaction of Hypernuclei.
%; neutrons were detected with a resolution of 13$\%$ at 10 MeV and 20$\%$ at 100 MeV. 
%All experimental details are given in \cite{plb701}.
We repeat the same procedure and define for each Hypernucleus the ratio $R_{1}$ as:
\begin{equation}
R_{1} \equiv  \ \frac{N_{n}(E_{p} \leq(\mu_{1} - \mathrm{20\ MeV}), cos \theta(np) \geq -0.8)}{N_{p}(E_{p}> \mu_{1})}
\label{R1def}
\end{equation}
where $N_{n}(E_{p} \leq(\mu_{1} - \mathrm{20\ MeV}), cos \theta(np) \geq -0.8)$ is the number of neutrons 
in coincidence with a proton of energy lower than $\mu_{1}-20$ MeV and forming an angle with the proton direction such as $cos \theta(np) \geq-0.8$, while $N_{p}$ is the number of protons with energy larger than 
$\mu_{1}$ (blue areas in Figure~\ref{pspectra}). These events \cite{plb701} should correspond mainly to the process (\ref{gamma2}) plus a not negligible contribution due to FSI. 
\begin{figure}[h]
\begin{center}
\includegraphics[width=90mm]{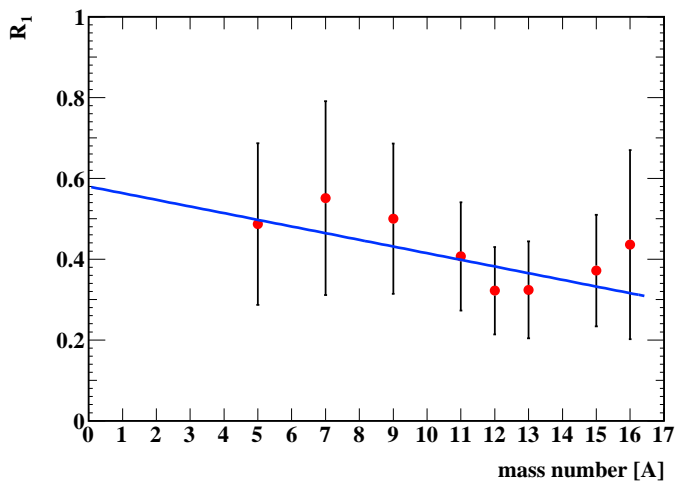}
\caption{(Color online) $R_{1}= N_{n}(E_{p} \leq(\mu_{1} - \mathrm{20\ MeV}), cos \theta(np) \geq -0.8)/N{p}(E_{p}> \mu_{1})$ values  as a function of $A$ for 
 ${\mathrm{^{5}_{\Lambda}He}}$,  ${\mathrm{^{7}_{\Lambda}Li}}$, ${\mathrm{^{9}_{\Lambda}Be}}$,  ${\mathrm{^{11}_{\Lambda}B}}$,  ${\mathrm{^{12}_{\Lambda}C}}$,  ${\mathrm{^{13}_{\Lambda}C}}$,  ${\mathrm{^{15}_{\Lambda}N}}$ and ${\mathrm{^{16}_{\Lambda}O}}$ from the present analysis. 
The blue line is a linear fit to the data; see text for more details.} 
\label{figRn}
\end{center}
\end{figure}

In Figure~\ref{figRn} the new experimental values of $R_{1}$ for each Hypernucleus are plotted as a function of $A$. By a simple linear fit to $(a+bA)$ (blue line in Figure~\ref{figRn}) we find the values 
$a=0.58\pm0.23$, $b=-0.017\pm0.090$ with $\chi^{2}$/ndf = 1.045/6 and then, following the 
approximations adopted in \cite{plb701}:

\begin{equation}                                                       
 \frac{\Gamma_{2N}}{\Gamma_{p}}=\frac{[R_{1}(A)-bA]}{1.6}= \frac{a}{1.6}=0.36\pm {0.14_{stat}}^{+0.05_{sys}}_{{-{0.04_{sys}}}}~.
\label{g2_n}
\end{equation}

Furthermore
\begin{equation}
\Gamma_{2N}/\Gamma_{NMWD} = 0.20 \pm {0.08_{stat}}^{+0.04_{sys}}_{{-{0.03_{sys}}}}~.
\label{res_n}
\end{equation}
The new estimations (\ref{g2_n}) and
 (\ref{res_n}) agree well with the previous ones \cite{plb701} ($\Gamma_{2N}/\Gamma_{p}=0.39\pm {0.16_{stat}}^{+0.04_{sys}}_{{-{0.03_{sys}}}}$, $\Gamma_{2N}/\Gamma_{NMWD}=0.21\pm {0.07_{stat}}^{+0.03_{sys}}_{{-{0.02_{sys}}}}$) and are in agreement with recent theoretical predictions \cite{bauer2} and the result from KEK \cite{kim} 
($\Gamma_{2N}/\Gamma_{NMWD}=0.29\pm0.13$). 
The systematic error quoted in (\ref{g2_n}) contains also the contribution, $\pm0.015$, due to the maximum difference obtained considering  
the fits 
from 60 MeV, $\Gamma_{2N}/\Gamma_{p}=0.37\pm {0.14_{stat}}^{+0.04_{sys}}_{{-{0.03_{sys}}}}$, and 80 MeV, $\Gamma_{2N}/\Gamma_{p}=0.39\pm {0.16_{stat}}^{+0.04_{sys}}_{{-{0.03_{sys}}}}$ \cite{plb701} too; 
in (\ref{res_n}) the systematic error is calculated by propagating the previous one. 

We finally recall that the analysis of the neutron-proton coincidences allowed us to find three candidate events for the $2N$ decay with full reconstruction of the final state particles $(n,n,p)$ kinematics \cite{nnp}.

\section{First determination of $\Gamma_{p}/\Gamma_{\Lambda}$ for eight hypernuclei $(A=5\div16)$ }

The previous studies \cite{plb685,plb701} demonstrated that the higher energy part of the 
bump observed around 80 MeV for all the examined Hypernuclei is due to the decay (\ref{gammap}), even 
though significantly distorted by FSI. 
In order to quantify $\Gamma_{2N}/\Gamma_{p}$, the important contribution due to FSI was parametrized by resorting to the measurement of relative quantities, the ratios $R$ and  $R_{1}$. 

On the contrary, in order to deduce the absolute values of $\Gamma_{p}/\Gamma_{\Lambda}$ from the measured spectra it is necessary to calculate as accurately as possible the effective influence of the FSI effect. 
%As anticipated in Section~1, FSI strongly affects the proton spectra. 
%FSI affect heavily the proton spectra and they must be determined accurately to infer correctly $\Gamma_{p}/\Gamma_{\Lambda}$. 
More in detail:
\begin{itemize}
\item[{\it a)}] the real number of primary protons due to decays (\ref{gammap}) and (\ref{gamma2}) is decreased due to FSI suffered by the proton;
\item[{\it b)}]	there is an increase of the number of protons due not only to FSI of protons at higher energy in the spectrum, but also to FSI of higher energy neutrons from (\ref{gamman}); 
\item[{\it c)}] quantum-mechanical interference effects may occur among protons of the same energy from the different sources (primary from (\ref{gammap}) and (\ref{gamma2}), secondary from FSI).
\end{itemize}
All these effects may be evaluated by appropriate and precise INC calculations, as done in \cite{garb1} and in \cite{bauer}. 

We try to evaluate the effect of the FSI on our spectra without using INC calculations but exploiting only  experimental data and simple hypotheses. If we consider the portions of the spectra above the $\mu_{1}$ values  (blue areas in Figure~1), the importance of the effect $b)$ may be safely neglected, following \cite{garb1}. The contribution of the decay (\ref{gamma2}) above 70 MeV is not larger than 5$\%$ of 
$\Gamma_{NMWD}$ \cite{garb1}, and, considering our determination (\ref{g2_n}), the total amount of primary protons from 
(\ref{gamma2}) would not be larger than 2$\%$ of those from (\ref{gammap}). Then also the interference effect $c)$ may be neglected.

We parametrize then the effect $a)$ by means of the following relationship:
\begin{equation}
\frac{\Gamma_{p}}{\Gamma_{\Lambda}} =  \frac{\Gamma_{T}}{\Gamma_{\Lambda}}\ BR(p)  = \frac{\Gamma_{T}}{\Gamma_{\Lambda}} 
\frac{2(N_{p} - N_{2N}) + \alpha(N_{p} - N_{2N})}{N_{hyp}}  
\label{alpha}
\end{equation}
where $BR(p)$ is the branching ratio of (\ref{gammap}), $N_{p}$ is the number of protons in the higher energy half part of the fitting Gaussian, $N_{2N}$ 
the number of protons from (\ref{gamma2}) (about 2$\%$), $N_{hyp}$ the number of  produced Hypernuclei, 
the factor 2 takes into account the total area of the Gaussians and $\alpha$ is a coefficient to be determined, which accounts for the number of protons moved below $\mu_{1}$ due to FSI. More precisely $\alpha/(2+\alpha)$ is the fraction of protons affected by FSI.

To calculate $\alpha$ for the considered Hypernuclei, $\Gamma_{p}/\Gamma_{\Lambda}$ values for 
$^{5}_{\Lambda}$He and $^{12}_{\Lambda}$C are considered and a linear scaling law with $A$ is 
assumed for the FSI contribution, and consequently for $\alpha$. 
$\Gamma_{p}/\Gamma_{\Lambda}$ for $^{5}_{\Lambda}$He and $^{12}_{\Lambda}$C can be evaluated from 
(\ref{gammatot2}), explicitely:
\begin{equation}
\frac{\Gamma_{T}}{\Gamma_{\Lambda}} = \frac{\Gamma_{\pi^{-}}}{\Gamma_{\Lambda}} + \frac{\Gamma_{\pi^{0}}}{\Gamma_{\Lambda}} + \frac{\Gamma_{p}}{\Gamma_{\Lambda}} +
\frac{\Gamma_{n}}{\Gamma_{p}} \cdot \frac{\Gamma_{p}}{\Gamma_{\Lambda}} +
\frac{\Gamma_{2N}}{\Gamma_{p}} \cdot \frac{\Gamma_{p}}{\Gamma_{\Lambda}}~,
\label{explicit}
\end{equation}
by means of the value of $\Gamma_{2N}/\Gamma_{p}$ given by (\ref{g2_n}) and other experimental values existing in the literature. 
More precisely for $^{5}_{\Lambda}$He, by substituting the experimental values of $\Gamma_{T}/\Gamma_{\Lambda}=0.96\pm0.03$ (w.~a.~of \cite{szym,kame}), $\Gamma_{n}/\Gamma_{p}= 0.45\pm0.11\pm0.03$ \cite{kang}, 
$\Gamma_{2N}/\Gamma_{p}= 0.36\pm {0.14_{stat}}^{+0.05_{sys}}_{{-{0.04_{sys}}}}$ (\ref{g2_n}), $\Gamma_{\pi^{-}}/\Gamma_{\Lambda}= 0.34\pm0.02$ (w.a. of  \cite{szym,kame,plb681}),  $\Gamma_{\pi^{0}}/\Gamma_{\Lambda}= 0.20\pm0.01$ (w.a. of  \cite{szym,okada}), we obtain $\Gamma_{p}/\Gamma_{\Lambda}= 0.22\pm0.03$, to be compared with 0.21$\pm$0.07 given in \cite{szym}. 

For $^{12}_{\Lambda}$C we use $\Gamma_{T}/\Gamma_{\Lambda}= 1.22\pm0.04$ (w.a. of  \cite{kame,park}), $\Gamma_{n}/\Gamma_{p}=0.51\pm0.13\pm0.05$ \cite{kim}, 
$\Gamma_{2N}/\Gamma_{p}= 0.36\pm {0.14_{stat}}^{+0.05_{sys}}_{{-{0.04_{sys}}}}$ (\ref{g2_n}), $\Gamma_{\pi^{-}}/\Gamma_{\Lambda}=0.12\pm0.01$ (w.a. of \cite{szym,noumi,sato,bhang}),  $\Gamma_{\pi^{0}}/\Gamma_{\Lambda}=0.17\pm0.01$ (w.a. of \cite{okada,saka}), and we obtain $\Gamma_{p}/\Gamma_{\Lambda}= 0.49\pm0.06$, to be compared with the values 0.31$\pm$0.07 given 
in \cite{noumi} and 0.45$\pm$0.10 given in \cite{kim,bhang}.

For sake of clarity, it must be observed that the available experimental determinations of 
$\Gamma_{p}/\Gamma_{\Lambda}$ for $^{5}_{\Lambda}$He \cite{szym} and $^{12}_{\Lambda}$C 
\cite{noumi,kim,bhang} were not used directly in (\ref{alpha}) to calculate the FSI correction factor $\alpha$:  those values, in fact, were obtained treating FSI with the help of INC calculations or simulations and could increase the systematic errors in our FSI effect evaluation.
%; on the contrary, the method we apply relates to direct measurements of decay widths and lifetimes, to $\Gamma_{n}/\Gamma_{p}$'s obtained in $np$ coincidence measurements 
%and to $\Gamma_{2N}/\Gamma_{p}$ based on systematic measurements on $p$-shell hypernuclei. 

With the indirect values of $\Gamma_{p}/\Gamma_{\Lambda}$ for $^{5}_{\Lambda}$He and $^{12}_{\Lambda}$C  we may obtain from (\ref{alpha}) two evaluations for $\alpha$, by using the above reported values of $\Gamma_{T}/\Gamma_{\Lambda}$ and the experimental values of $N_{p}$ (70 MeV fit) and $N_{hyp}$.  \\
We find $\alpha_{5}$($^{5}_{\Lambda}$He)=1.15$\pm$0.26 for $^{5}_{\Lambda}$He (indicated as subscript) from $^{5}_{\Lambda}$He measurements (indicated between parentheses) and $\alpha_{12}$($^{12}_{\Lambda}$C)=2.48$\pm$0.46 for $^{12}_{\Lambda}$C from $^{12}_{\Lambda}$C measurements. By assuming that $\alpha$ scales linearly with $A$, it is straightforward to obtain the crossed evaluations: 
$\alpha_{5}$($^{12}_{\Lambda}$C)=1.04$\pm$0.19 and $\alpha_{12}$($^{5}_{\Lambda}$He)=2.77$\pm$0.63. 
The w.a. of the two evaluations are $\overline{\alpha}_{5}$=1.08$\pm$0.16 and $\overline{\alpha}_{12}$=2.58$\pm$0.37. We adopt finally the general expression for $\alpha_{A}$:

\begin{equation}
\alpha_{A}= \frac{\overline{\alpha}_{5}}{5} \cdot A = \frac{\overline{\alpha}_{12}}{12} \cdot A = (0.215\pm 0.031) \cdot A
%\alpha_{A}= \frac{\overline{\alpha}_{5}}{5} \cdot A = \frac{\overline{\alpha}_{12}}{12} \cdot A = (0.215\pm 0.031_{stat}\pm0.013_{sys}) \cdot A
\label{alpha_a}
\end{equation}
where the statistical error comes from the errors on the quantities used to evaluate $\alpha$. 
A  systematic error can be evaluated by taking into account the difference between $\alpha_{5}$($^{5}_{\Lambda}$He) and 
$\alpha_{5}$($^{12}_{\Lambda}$C) for $^{5}_{\Lambda}$He and between $\alpha_{12}$($^{12}_{\Lambda}$C) and 
$\alpha_{12}$($^{5}_{\Lambda}$He) for $^{12}_{\Lambda}$C: this error amounts to $6\%$. 
It is also worth to observe that in \cite{plb685} and \cite{plb701} the assumption $\Gamma_{2N}\simeq\Gamma_{np}$ was made, 
which gives a systematic underestimation of $\Gamma_{2N}/\Gamma_{p}$ of $\sim$16$\%$ (much smaller than the experimental errors): 
we remind, indeed, that following \cite{bauer09} $\Gamma_{np} : \Gamma_{pp} : \Gamma{nn} = 0.83 : 0.12: 0.04$. 
If this systematic effect is taken into account in the calculation of $\alpha$ from (\ref{explicit}) and (\ref{alpha}), a decrease of $(9\div5) \%$ arises 
which gives a further systematic error on $\alpha(A=5\div16)$, for a total of $(10\div7)\%$. 

We remark that the hypothesis that FSI effects are to a first approximation proportional to $A$ was already adopted in \cite{plb685,plb701}. With (\ref{alpha_a}) we find that 35$\%$ of the primary protons from NMWD  are lost (moved below $\mu_{1}$) for FSI in $^{5}_{\Lambda}$He and 63$\%$ in $^{16}_{\Lambda}$O. 

We are thus able to determine with the equation (\ref{alpha}) the values of $\Gamma_{p}/\Gamma_{\Lambda}$ for $^{5}_{\Lambda}$He and all studied $p$-shell Hypernuclei. 
\begin{table}[h] 
\begin{center} 
\begin{tabular}{|c|c|c|c|c|c|} 
\hline 
 & $\Gamma_{T}/\Gamma_{\Lambda}$ & $\alpha_{A}$ & $\Gamma_{p}/\Gamma_{\Lambda}$ & $\Gamma_{p}/\Gamma_{\Lambda}$ & $\Gamma_{p}/\Gamma_{\Lambda}$ \\
& & & this work & previous works & \cite{motoba} \\ 
 \hline 
${\mathrm{^{5}_{\Lambda}He}}$ &0.96$\pm$0.03  & 1.08$\pm$0.16 & 0.22$\pm$0.05 &  0.21$\pm$0.07 \cite{szym} & 0.237\\ 
& \cite{szym,kame} & & & & \\
\hline
${\mathrm{^{7}_{\Lambda}Li}}$ & 1.12$\pm$0.12 & 1.51$\pm$0.22 & 0.28$\pm$0.07 &  & 0.297 \\ 
\hline
${\mathrm{^{9}_{\Lambda}Be}}$ & 1.15$\pm$0.13 & 1.94$\pm$0.28 & 0.30$\pm$0.07 &  & 0.401 \\ 
\hline 
${\mathrm{^{11}_{\Lambda}B}}$ & 1.28$\pm$0.10 \cite{noumi} & 2.37$\pm$0.34 & 0.47$\pm$0.11 & 
0.30$\pm$0.07 \cite{noumi} & 0.444 \\ 
\hline 
${\mathrm{^{12}_{\Lambda}C}}$ & 1.242$\pm$0.042 & 2.58$\pm$0.37 & 0.65$\pm$0.19 & 0.31$\pm$0.07 \cite{noumi} & 0.535 \\
& \cite{kame,park} & & & 0.45$\pm$0.10 \cite{bhang} &  \\
\hline 
${\mathrm{^{13}_{\Lambda}C}}$ & 1.21$\pm$0.16 & 2.80$\pm$0.40 & 0.60$\pm$0.14 & & 0.495 \\  
\hline 
${\mathrm{^{15}_{\Lambda}N}}$ & 1.26$\pm$0.18 & 3.23$\pm$0.47 & 0.49$\pm$0.11 & & 0.555 \\  
\hline 
${\mathrm{^{16}_{\Lambda}O}}$ & 1.28$\pm$0.19 & 3.44$\pm$0.50 & 0.44$\pm$0.12 & & 0.586 \\  
\hline 
\end{tabular} 
\caption{First column: hypernucleus; second column: total decay width $\Gamma_{T}$ in units of the free $\Lambda$ decay width $\Gamma_{\Lambda}$; third column: $\alpha$ factor; fourth column: present evaluation of 
$\Gamma_{p}/\Gamma_{\Lambda}$; fifth column: previous measurements; sixth column: recent theoretical calculation of $\Gamma_{p}/\Gamma_{\Lambda}$ \cite{motoba}. 
} 
\label{table2} 
\end{center} 
\end{table}
They are given in Table~\ref{table2}, which reports also the experimental values of 
$\Gamma_{T}/\Gamma_{\Lambda}$ we used, when available; 
for the other Hypernuclei we adopt the parametrization $\Gamma_{T}/\Gamma_{\Lambda}(A) = (0.990\pm0.094) + (0.018\pm0.010)\cdot A$ proposed in \cite{plb681}. 
The errors on $\Gamma_{p}/\Gamma_{\Lambda}$ are calculated by considering statistical errors for $N_{p}$ and $N_{hyp}$,  
the errors reported in column two for $\Gamma_{T}/\Gamma_{\Lambda}(A)$ and the statistical error for $\alpha_{A}$, 
which is the largest one (15$\%$). 
%This error includes possible systematic contributions, like those due the determination of the values of $\mu_{1}$.
Table~\ref{table2} reports, in the sixth column, the theoretical values of $\Gamma_{p}/\Gamma_{\Lambda}$ calculated recently in \cite{motoba}.

It is interesting to note that the new evaluation, following the determination of $\alpha_{A}$ given by (\ref{alpha_a}), and the former indirect calculation for both $^{5}_{\Lambda}$He and $^{12}_{\Lambda}$C 
from (\ref{explicit}) are compatible within the error: thus it appears that the method used to evaluate 
$\alpha_{A}$ does not introduce any systematic error in the $\Gamma_{p}/\Gamma_{\Lambda}$ value. 
On the other hand, a systematic error on $\Gamma_{p}/\Gamma_{\Lambda}$ can be estimated by repeating all the previous procedure with 
the fits from 60 MeV and from 80 MeV: it corresponds to $5\div6 \%$ for ${\mathrm{^{5}_{\Lambda}He}}$, ${\mathrm{^{7}_{\Lambda}Li}}$, ${\mathrm{^{9}_{\Lambda}Be}}$, ${\mathrm{^{16}_{\Lambda}O}}$ and $9\div10 \%$ for ${\mathrm{^{11}_{\Lambda}B}}$, ${\mathrm{^{12}_{\Lambda}C}}$, ${\mathrm{^{13}_{\Lambda}C}}$ and ${\mathrm{^{15}_{\Lambda}N}}$. 
Another systematic error on $\Gamma_{p}/\Gamma_{\Lambda}$ can be obtained from the systematic error on $\alpha(A)$: it amounts to 
$(3\div3.5\%)$ only; the total systematic error amounts to $6\div7 \%$ for ${\mathrm{^{5}_{\Lambda}He}}$, ${\mathrm{^{7}_{\Lambda}Li}}$, ${\mathrm{^{9}_{\Lambda}Be}}$, ${\mathrm{^{16}_{\Lambda}O}}$ and $10\div11\%$ for ${\mathrm{^{11}_{\Lambda}B}}$, ${\mathrm{^{12}_{\Lambda}C}}$, ${\mathrm{^{13}_{\Lambda}C}}$ and ${\mathrm{^{15}_{\Lambda}N}}$ and is small compared to the statistical one. 

\begin{figure}[hp]
\begin{center}
\includegraphics[width=100mm]{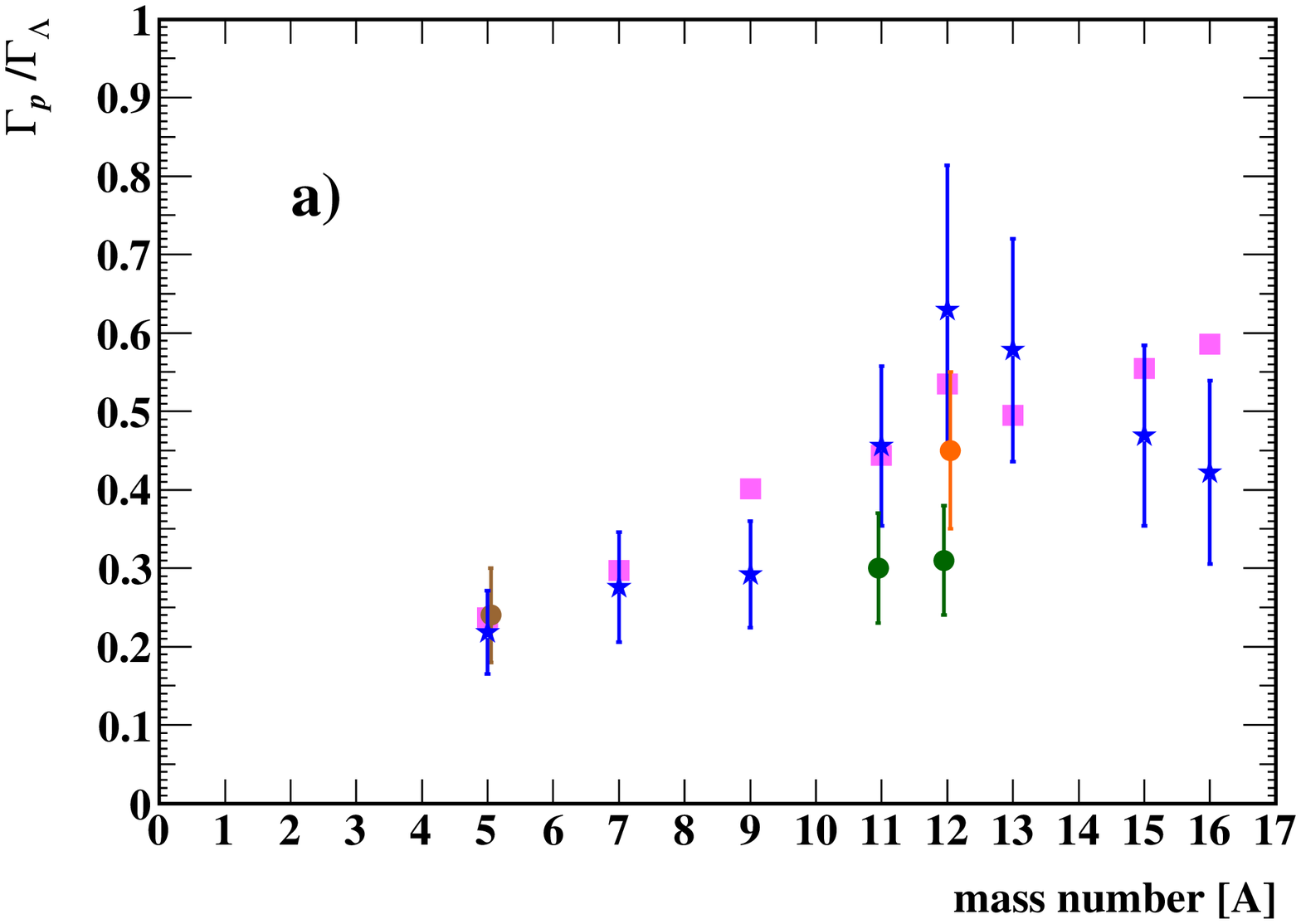}
\hspace{0.1mm}
\includegraphics[width=100mm]{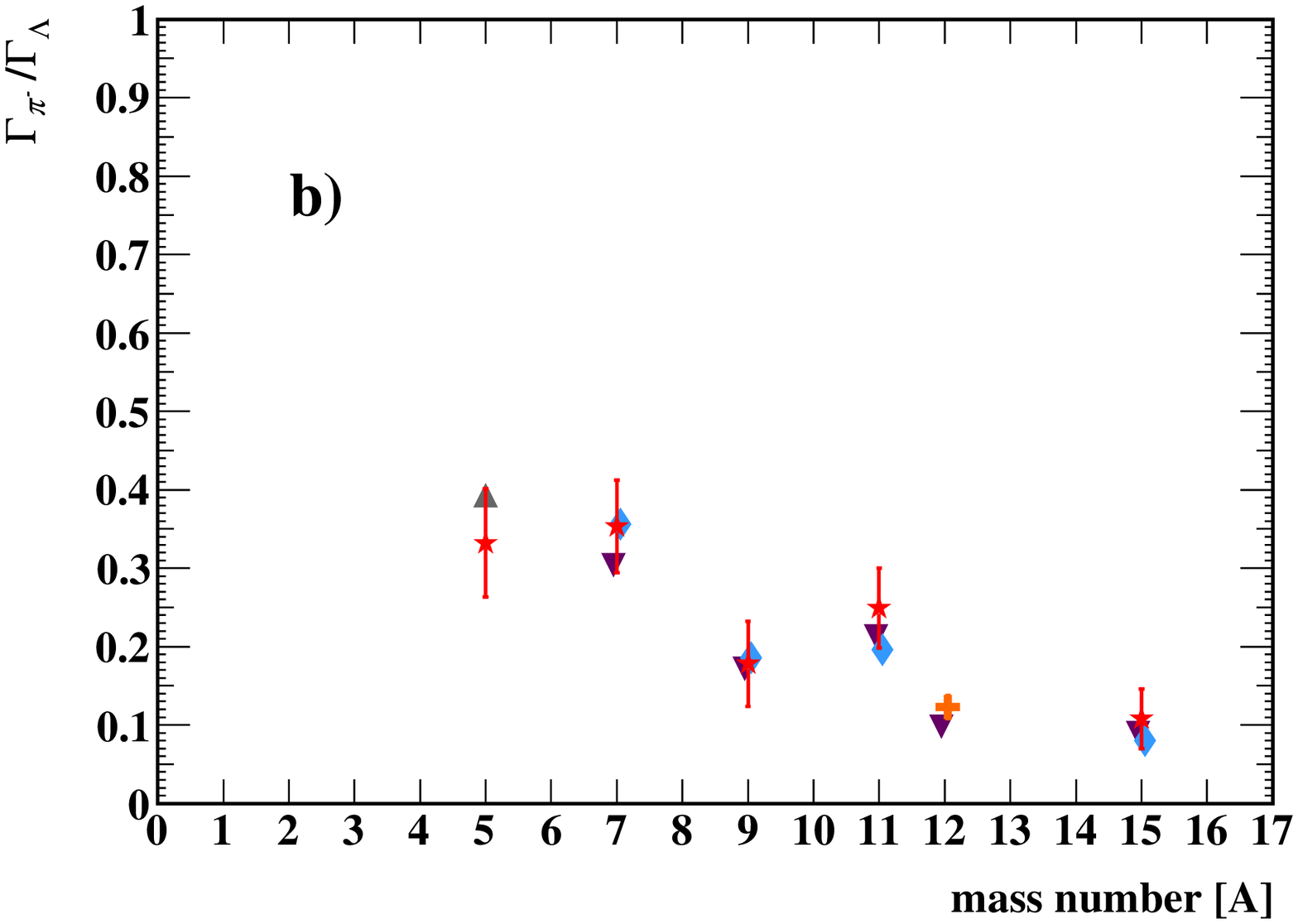}
\caption{(Color online) a) $\Gamma_{p}/\Gamma_{\Lambda}$ values as a function of $A$ for 
 ${\mathrm{^{5}_{\Lambda}He}}$,  ${\mathrm{^{7}_{\Lambda}Li}}$, ${\mathrm{^{9}_{\Lambda}Be}}$,  ${\mathrm{^{11}_{\Lambda}B}}$,  ${\mathrm{^{12}_{\Lambda}C}}$,  ${\mathrm{^{13}_{\Lambda}C}}$,  ${\mathrm{^{15}_{\Lambda}N}}$ and ${\mathrm{^{16}_{\Lambda}O}}$ from the present analysis (blue stars, errors from Table~\ref{table2}). Theoretical calculations of $\Gamma_{p}/\Gamma_{\Lambda}$ \cite{motoba} (violet squares) are also shown. $\Gamma_{p}/\Gamma_{\Lambda}$ from \cite{szym} for ${\mathrm{^{5}_{\Lambda}He}}$ (brown full circle), 
from \cite{noumi} for $^{11}_{\Lambda}$B and $^{12}_{\Lambda}$C (green full circles) 
and from \cite{bhang} for $^{12}_{\Lambda}$C (orange full circle) have also been plotted. 
b) $\Gamma_{\pi^{-}}/\Gamma_{\Lambda}$ values as a function of $A$  
for ${\mathrm{^{5}_{\Lambda}He}}$,  ${\mathrm{^{7}_{\Lambda}Li}}$, ${\mathrm{^{9}_{\Lambda}Be}}$, ${\mathrm{^{11}_{\Lambda}B}}$ and ${\mathrm{^{15}_{\Lambda}N}}$ (red stars)  from \cite{plb681}  
and for ${\mathrm{^{12}_{\Lambda}C}}$ (orange cross) from \cite{bhang}. Theoretical calculations of  
$\Gamma_{\pi^{-}}/\Gamma_{\Lambda}$ for ${\mathrm{^{5}_{\Lambda}He}}$ (gray up triangle) from \cite{motoba5He}, for $^{7}_{\Lambda}$Li, $^{9}_{\Lambda}$Be, $^{11}_{\Lambda}$B, $^{12}_{\Lambda}$C and $^{15}_{\Lambda}$N (down violet triangles) from \cite{motoba} and for ${\mathrm{^{5}_{\Lambda}He}}$,  ${\mathrm{^{7}_{\Lambda}Li}}$, ${\mathrm{^{9}_{\Lambda}Be}}$, ${\mathrm{^{11}_{\Lambda}B}}$ and ${\mathrm{^{15}_{\Lambda}N}}$ (cyan diamonds) from \cite{gal} are also reported.
}
\label{figGp}. 
\end{center}
\end{figure}
Figure~\ref{figGp}a) shows the comparison among the values from this experiment (blue stars), 
previous data \cite{szym,noumi,bhang} (brown full circle, green full circles and orange full circle 
respectively) and theoretical values \cite{motoba} (violet squares); in the figure statistical errors only are indicated on the present results 
to match with previous data.
A general agreement between our data and the theoretical ones 
is evident, even though the experimental errors are quite large, with the exception of 
$^{9}_{\Lambda}$Be, which is lower by  1.5 $\sigma$ and of $^{16}_{\Lambda}$O which is lower by 1.3 
$\sigma$. 
Figure~\ref{figGp}b) shows the experimental determinations of $\Gamma_{\pi^{-}}/\Gamma_{\Lambda}$ 
from \cite{plb681} (red stars) for ${\mathrm{^{5}_{\Lambda}He}}$,  ${\mathrm{^{7}_{\Lambda}Li}}$, ${\mathrm{^{9}_{\Lambda}Be}}$, ${\mathrm{^{11}_{\Lambda}B}}$ and ${\mathrm{^{15}_{\Lambda}N}}$ 
and from \cite{bhang} (orange cross) for ${\mathrm{^{12}_{\Lambda}C}}$; theoretical calculations of  
$\Gamma_{\pi^{-}}/\Gamma_{\Lambda}$ from \cite{motoba5He} (gray up triangle) for ${\mathrm{^{5}_{\Lambda}He}}$, from \cite{motoba} (down violet triangles) for $^{7}_{\Lambda}$Li, $^{9}_{\Lambda}$Be, $^{11}_{\Lambda}$B $^{12}_{\Lambda}$C and $^{15}_{\Lambda}$N and from 
\cite{gal} (cyan diamonds) for ${\mathrm{^{5}_{\Lambda}He}}$,  ${\mathrm{^{7}_{\Lambda}Li}}$, ${\mathrm{^{9}_{\Lambda}Be}}$, ${\mathrm{^{11}_{\Lambda}B}}$ and ${\mathrm{^{15}_{\Lambda}N}}$ 
are also indicated.
The purpose is to show the 
first experimental verification, at least for the $\Gamma$'s relative to charged particles, of the long-time advocated complementary behaviour of MWD and NMWD of Hypernuclei in the relevant $A$ range ($5\div16$). 

\section{Conclusions}
We have determined the partial decay widths of the one-proton induced NMWD from measured proton spectra for eight Hypernuclei ($A=5\div16$): it is the first systematic determination ever done for $p$-shell $\Lambda$-Hypernuclei. The measured values, though affected by errors ranging from 20$\%$ to 30$\%$, agree reasonably with those predicted by a recent precise theoretical calculation \cite{motoba}. 

To make a better comparison smaller experimental errors are necessary, at a 10$\%$ level or less, on both 
$\Gamma_{T}/\Gamma_{\Lambda}$ and $N_{p}$ ($N_{hyp}$) (see (\ref{alpha})). 
At present, the Laboratory equipped with the beams and detector’s arrays necessary for this kind of measurements is J-PARC. It is also necessary to develop a simple INC calculation, without 
two-nucleon induced contributions, to verify the validity of the hypothesis of linearity with $A$ of the FSI correction. 

With all these requirements satisfied, it could then be possible to try to face the problem of the experimental study of the  $\Lambda N \rightarrow NN$ Weak Interaction from the Hypernuclear data.
Indeed, a precise study of the NMWD of Hypernuclei will be the only way to get 
information on the four-baryon weak process $\Lambda N \rightarrow N N$ and to realize then the idea of 
using a nuclear system as a “Laboratory” for the study of interactions between elementary particles not 
otherwise accessible in vacuo.

\end{document}